\documentclass[twocolumn,prb,superscriptaddress,aps,prd,longbibliography]{revtex4-1}
\usepackage{graphicx}
\usepackage[cmex10]{amsmath}
\usepackage[utf8]{inputenc}
\usepackage{float}
\usepackage{amssymb}
\usepackage[T1]{fontenc}
\usepackage{hyperref}
\begin{document}

\title{Antiferromagnetic nano-oscillator in external magnetic fields}

\author{Jakub Ch\k{e}ciński}

\affiliation{AGH University of Science and Technology, Department of Electronics, Al. Mickiewicza 30, 30-059 Krak\'{o}w, Poland}
\affiliation{AGH University of Science and Technology, Faculty of Physics and Applied Computer Science, Al. Mickiewicza 30, 30-059 Krak\'{o}w, Poland}

\author{Marek Frankowski}

\affiliation{AGH University of Science and Technology, Department of Electronics, Al. Mickiewicza 30, 30-059 Krak\'{o}w, Poland}

\author{Tomasz Stobiecki}
\affiliation{AGH University of Science and Technology, Department of Electronics, Al. Mickiewicza 30, 30-059 Krak\'{o}w, Poland}
\affiliation{AGH University of Science and Technology, Faculty of Physics and Applied Computer Science, Al. Mickiewicza 30, 30-059 Krak\'{o}w, Poland}

\date{\today}

\begin {abstract}

We describe the dynamics of an antiferromagnetic nano-oscillator in an external magnetic field of any given time distribution. The oscillator is powered by a spin current originating from spin-orbit effects in a neighboring heavy metal layer, and is capable of emitting a THz signal in the presence of an additional easy-plane anisotropy. We derive an analytical formula describing the interaction between such a system and an external field, which can affect the output signal character. Interactions with magnetic pulses of different shapes, with a sinusoidal magnetic field and with a sequence of rapidly changing magnetic fields are discussed. We also perform numerical simulations based on the Landau-Lifshitz-Gilbert equation with spin-transfer torque effects to verify the obtained results and find a very good quantitative agreement between analytical and numerical predictions.

\end{abstract}

\maketitle

\section{Introduction}

Antiferromagnetic spintronics is a rapidly developing domain that can bring a lot of potential applications with respect to magnetic memories, magnonics or spin caloritronics \cite{jungwirth2016antiferromagnetic,cheng2016antiferromagnetic,wu2016antiferromagnetic}. Recently, it has been proposed to utilize an antiferromagnet in oscillators producing signal in a terahertz range \cite{gomonay2014spintronics,cheng2016terahertz,khymyn2017antiferromagnetic}. The working principle of such a device relies on a spin-orbit interaction taking place in a neighboring metallic layer and injecting a spin current into the antiferromagnet, where it exerts a spin-transfer torque that can sustain a stable oscillation in the presence of additional anisotropy \cite{cheng2016terahertz,khymyn2017antiferromagnetic}. By the virtue of spin pumping effects \cite{tserkovnyak2002enhanced,tserkovnyak2002spin,tserkovnyak2005nonlocal,nakayama2012geometry,gomonay2014spintronics}, this oscillation can produce an electric field that, depending on the type of antiferromagnet material used and on the powering current density, may be in the range of hundreds of GHz or THz. Such a feature is extremely interesting from the application point of view, since reliable signal processing at this frequency has been often identified as a challenging technological issue \cite{sirtori2002applied,tonouchi2007cutting,gulyaev2014generation,khymyn2017antiferromagnetic}.

The dynamics of antiferromagnetic oscillators is typically described by a system of coupled equations which can later be reduced to a single equation describing the Néel vector time evolution \cite{cheng2016terahertz,khymyn2017antiferromagnetic}. Using such an approach, multiple physical quantities, including the threshold current, the oscillation frequency or the linewidth, can be derived. We propose to expand this framework by including the effects of an interaction between a working oscillator powered by a spin current and an external magnetic field. Similar interactions between an antiferromagnetic material and an external field have been already described in the context of magnetic field rapid pulse switching \cite{jungwirth2016antiferromagnetic,kampfrath2011coherent,
wienholdt2012thz,kim2014coherently,tao2016switching,kim2017field}, but not in the context of an oscillator emitting a THz signal. A generalized solution to the Néel vector dynamics equation and a formula describing the response to any given external field will be derived. A number of specific cases of interactions with an external field that are particularly interesting from either fundamental or application point of view will also be discussed. We believe that, although controlling the magnetization with an external magnetic field is generally considered undesirable in the modern spintronics, our findings may contribute into the understanding of the antiferromagnetic oscillator dynamics as well as assist in the future development of its applications. 

\begin{figure}
\centering
\includegraphics[trim=0cm 0cm 0cm 1cm, width=8.5cm]{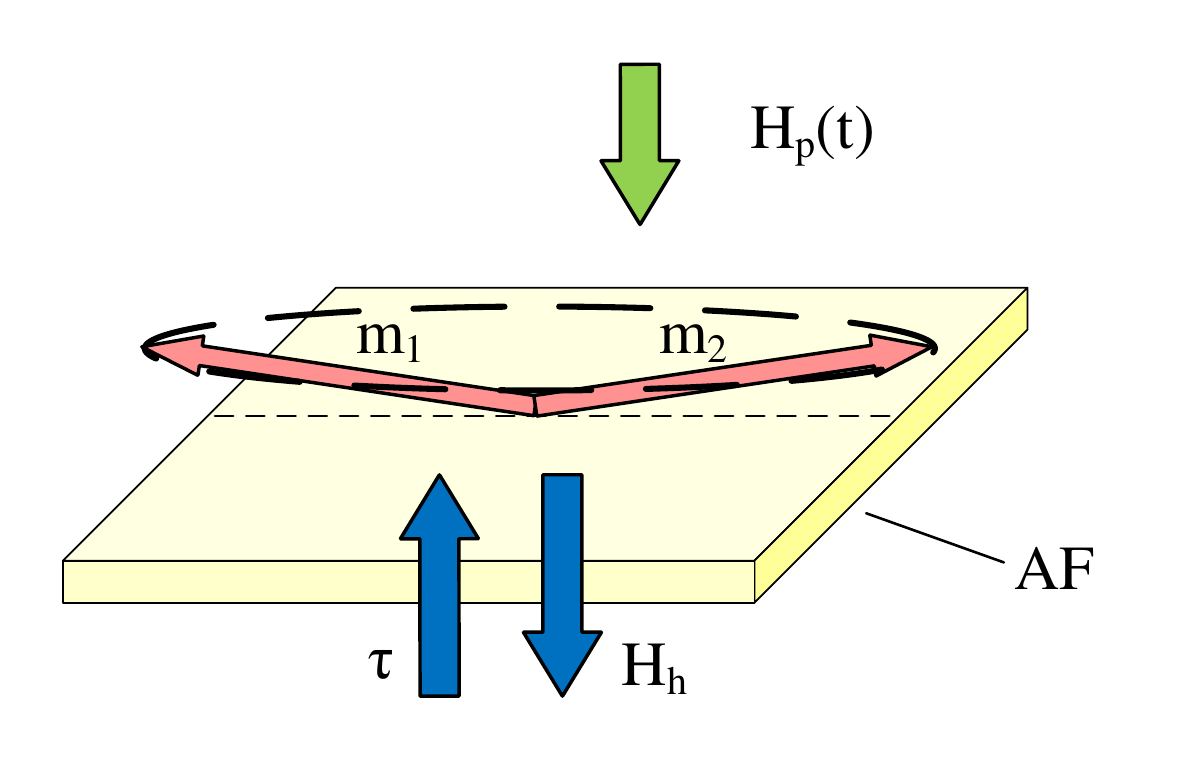}
\caption{Schematic picture of the antiferromagnetic nano-oscillator interacting with external magnetic field. Bright red arrows represent the antiferromagnetic magnetization vectors $\vec{m}_1$ and $\vec{m}_2$, blue arrows represent the influence of hard axis anisotropy field $\vec{H}_h$ and of the spin torque $\vec{\tau}$ (both constant in time). Green arrow represents the external magnetic field $\vec{H}_p$ (varying in time) which is either parallel or anti-parallel to the anisotropy field.}
\label{fig1}
\end{figure}

A schematic representation of the considered system can be seen in Fig \ref{fig1}. A spin-polarized current is injected into a thin antiferromagnetic layer, exerting a torque in the direction opposite to the direction of the hard axis anisotropy field and leading to an occurrence of magnetization vectors joint precession. The characteristic frequency of this precession is determined mostly by the antiferromagnetic exchange field amplitude. In a typical antiferromagnetic material, the exchange field is of the order of magnitude of $\mathrm{10^3}$ T and the obtained oscillations are thus close to the THz range. In addition to this picture, we introduce a time-dependent external field vector $\vec{H}_p(t)$, which will further alter the dynamics of the system. Since fields lying in the antiferromagnet easy plane are not expected to influence the oscillations greatly, we limit the direction of the external field to the hard axis direction.

\section{General antiferromagnetic oscillator dynamics}
\label{sec2}

We consider an antiferromagnetic oscillator described by two magnetization vectors $\vec{m_1}$ and $\vec{m_2}$ with dynamics governed by the following Landau-Lifshitz-Gilbert-Slonczewski (LLGS) equations: \cite{landau1935theory,gilbert1955lagrangian,gilbert2004phenomenological,slonczewski1996current,
berger1996emission,gurevich1996magnetization,gomonay2014spintronics,khymyn2017antiferromagnetic}

\begin{equation}
\begin{split}
\frac{d\vec{m}_1}{dt}=-\gamma\left(\vec{m}_1 \times \vec{H}_1 + \alpha \vec{m}_1 \times (\vec{m}_1\times \vec{H}_1) \right) + \\
\tau\vec{m}_1 \times (\vec{m}_1\times \vec{p}),
\label{LLG1}
\end{split}
\end{equation}

\begin{equation}
\begin{split}
\frac{d\vec{m}_2}{dt}=-\gamma\left(\vec{m}_2 \times \vec{H}_2 + \alpha \vec{m}_2 \times (\vec{m}_2\times \vec{H}_2) \right) + \\
\tau\vec{m}_2 \times (\vec{m}_2\times \vec{p}),
\label{LLG2}
\end{split}
\end{equation}

where $\gamma$ is the electron gyromagnetic ratio equal to 176.1 GHz/T, $\alpha$ is the damping coefficient, $\vec{p}$ is a unit vector along the spin current polarization, $\tau$ is the spin-transfer torque expressed in frequency units and $\vec{H_1},\vec{H_2}$ are the effective fields for both vectors. In our case, there are three main contributions to the effective field: the antiferromagnetic exchange field $\vec{H_{ex}}$, the hard axis anisotropy field $\vec{H_{h}}$ and the external field $\vec{H_{p}}$, which leads to the equations:

\begin{equation}
\begin{split}
\vec{H}_1 = -\frac{1}{2}H_{ex}\vec{m}_2-H_h\vec{n}_h(\vec{n}_h\cdot\vec{m}_1)+\vec{H}_{p},
\end{split}
\end{equation}

\begin{equation}
\begin{split}
\vec{H}_2 = -\frac{1}{2}H_{ex}\vec{m}_1-H_h\vec{n}_h(\vec{n}_h\cdot\vec{m}_2)+\vec{H}_{p},
\end{split}
\end{equation}

where $\vec{n_h}$ is a unit vector along the hard anisotropy axis. In practice, we limit our investigation only to the cases where $\vec{H_{p}}$ is parallel or anti-parallel to $\vec{n_h}$. Following the procedure described in Ref. \cite{khymyn2017antiferromagnetic} and assuming no external field present, equations \ref{LLG1} and \ref{LLG2} can be rewritten as differential equations for spherical coordinates $\phi$ and $\theta$ of the antiferromagnetic Néel vector \cite{poole2004encyclopedic,ivanov2014spin} $\vec{l} = (\vec{m_1}-\vec{m_2})/2$:

\begin{equation}
\begin{split}
\frac{\dot{\theta}\dot{\phi}}{\omega_{ex}}sin(2\theta)+sin^2\theta\left(\frac{\ddot{\phi}}{\omega_{ex}}+\alpha\dot{\phi}+\tau\right)=0,
\end{split}
\end{equation}

\begin{equation}
\begin{split}
\frac{\ddot{\theta}}{\omega_{ex}}+\alpha\dot{\theta}-sin\theta cos\theta \left( \frac{(\dot{\phi})^2}{\omega_{ex}}+\omega_H\right)=0,
\end{split}
\end{equation}

where $\omega_{ex} = \gamma H_{ex}$ and $\omega_H (t) = \gamma H_{h}$. The equations above can be solved for the special case of $\theta$ = $\pi$/2 (rotation taking place in the easy antiferromagnet plane only), leading to a single equation for $\phi$:

\begin{equation}
\begin{split}
\frac{\ddot{\phi}}{\omega_{ex}}+\alpha\dot{\phi}+\tau=0.
\label{phieq}
\end{split}
\end{equation}

It can be seen immediately that $\dot{\phi} = -\tau/\alpha$ satisfies the equation above. In the case of external magnetic field being present, additional terms will arise in the Lagrangian describing the antiferromagnet dynamics \cite{ivanov2005mesoscopic,gomonaui2008distinctive,ivanov2014spin,khymyn2016transformation}, leading to a modified version of the equation for $\phi$ at $\theta = \pi/2$:
\begin{equation}
\begin{split}
\frac{\ddot{\phi}}{\omega_{ex}}+\alpha\dot{\phi}+\tau+\frac{\gamma\frac{dH_p}{dt}}{\omega_{ex}}=0.
\label{phieq_new}
\end{split}
\end{equation}

with a new factor in the solution describing the interaction between the antiferromagnetic oscillator and the external magnetic field pointing along the system hard axis:

\begin{equation}
\begin{split}
\dot{\phi}(t)=-\tau/\alpha+\gamma e^{-\lambda t}\int_{-\infty}^t \left(\frac{dH_{p}}{dt'}\right) e^{\lambda t'} dt',
\label{general-solution}
\end{split}
\end{equation}
where $\lambda = \omega_{ex}\alpha$. The time evolution of the $\dot{\phi}$ derivative (and, therefore, of the output electric field) is fully determined by the history of external magnetic field changes $dH_{p}/dt'$ and the parameters of the system. \\

\begin{figure}
\centering
\includegraphics[width=9cm]{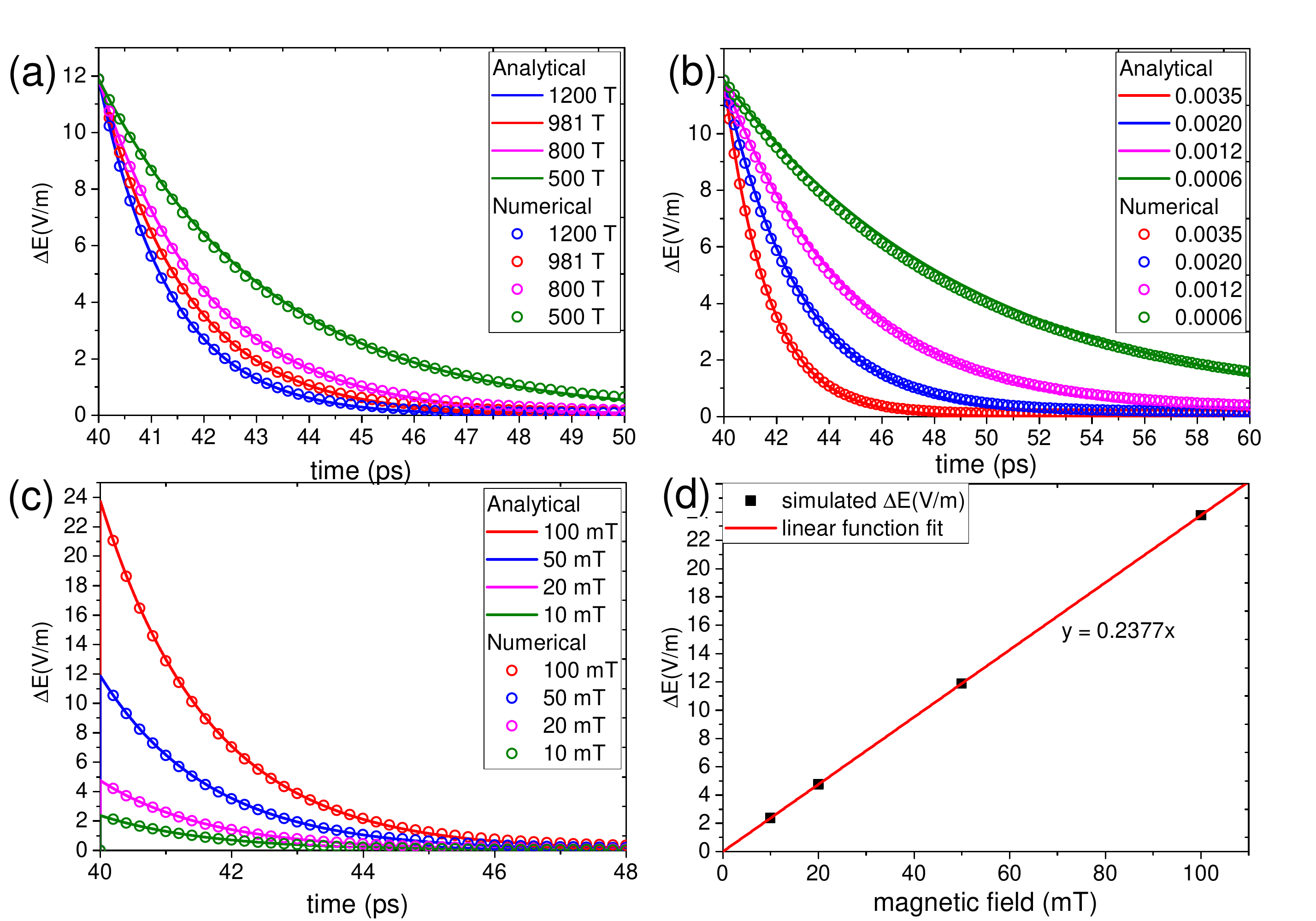}
\caption{Comparison of the electric field outputs obtained from numerical simulations (empty circles) of eq. \ref{LLG1} and \ref{LLG2} and from the analytical solution (solid lines) \ref{general-solution} for (a) different values of $\mu_0H_{ex}$, (b) different values of $\alpha$, (c) different values of $\mu_0H_p$. If not specified otherwise, $\mu_0H_{ex}$ was equal to 981 T, $\alpha$ was equal to 0.0035 and $\mu_0H_p$ to 50 mT. Figure (d) shows the value of electric field change as a function of the external magnetic field step amplitude together with a linear fit.}
\label{fig2}
\end{figure}

To gain an additional insight into the antiferromagnetic oscillator dynamics, we conducted numerical simulations of a system governed by equations \ref{LLG1} and \ref{LLG2} in the presence of a magnetic field step function. The parameters of the oscillator, if not specified otherwise, were similar to these in the Ref. \cite{khymyn2017antiferromagnetic} (i.e., an antiferromagnetic insulator characterized by Néel temperature significantly above the room temperature, such as NiO \cite{lewis1973thermal}) and were as following: $\alpha$\nobreakspace =\nobreakspace 0.0035, $\mu_0H_{ex}$ = 981 T ($\omega_{ex}$ = 2$\mathrm{\pi}$ $\times$ 27.5\nobreakspace THz), $\mu_0H_h$\nobreakspace =\nobreakspace 1.57 T, $\tau$ = -10.87 GHz, $\mu_0H_p$ = 50 mT. The driving spin current was polarized along the antiferromagnet hard axis direction. The obtained magnetization record was translated into the $\dot{\phi}$ value and then into electric field output using a proportionality constant $\kappa$\nobreakspace =\nobreakspace 1.35e-9 (V/m)/($\mathrm{rad/s}$), following Ref. \cite{khymyn2017antiferromagnetic}. The simulation results can be seen in Fig. \ref{fig2} (empty circles) calculated for (a) different $\mu_0H_{ex}$ values, (b) different $\alpha$ values, (c) different field step amplitude $\mu_0H_p$ values. The analytical predictions given by eq. \ref{general-solution} are included in the Fig. \ref{fig2} (a)-(c) as solid lines. Additionally, the initial value of the output electric field (d) is shown as a function of the external field step amplitude together with a linear fit. The fitted slope value is 237.7 (V/m)/T, which corresponds to the product of gyromagnetic ratio $\gamma$ and proportionality constant $\kappa$ and thus agrees with equation \ref{general-solution} for a magnetic field step excitation.

\section{Interactions with specific magnetic fields}

In this section, we will use the equation \ref{general-solution} to derive the antiferromagnetic oscillator response to a number of specific magnetic field excitations. We will present both the analytical calculations and the results of the numerical calculations conducted for an example system described in the previous section. 

\subsection{Magnetic field pulse}

One of the most commonly encountered types of magnetic field excitation is a single magnetic pulse characterized by a rectangular, trapezoidal or Gaussian-like shape. Since an antiferromagnetic oscillator does not respond to a constant magnetic field level, but only to the field derivative, the response to a rectangular pulse is described by a simple superposition of two responses to a single step (Fig. \ref{fig2}), with their values having opposite signs.

To describe the interaction with a trapezoidal magnetic field pulse, a response to a linearly changing field has to be derived first. By assuming $H_p(t') = \beta t'$ during a certain time period $t_A < t < t_B$ and field derivative equal to zero for other $t$ values, we can obtain:

\begin{equation}
\begin{split}
\dot{\phi}(t)=-\tau/\alpha+\gamma e^{-\lambda t}\int_{t_A}^{t_B} \beta e^{\lambda t'} dt' = \\
=-\tau/\alpha+\frac{\gamma \beta}{\lambda}\left(e^{\lambda(t_B-t)}-e^{\lambda(t_A-t)}\right).
\end{split}
\label{eq12}
\end{equation}

The full response to a trapezoidal pulse is a simple superposition of two responses to a linearly changing field, each with their respective $\beta$ value describing the field slope. For $t_A < t < t_B$, the upper bound of the integral $t_B$ is replaced by $t$. 

The response to a Gaussian-shaped field pulse characterized by an amplitude $H_0$, center position $t_\mu$ and variance $t_\sigma^2$ is given by:

\begin{equation}
\begin{split}
\dot{\phi}(t)=-\tau/\alpha+\gamma H_0 e^{-\lambda t}\int_{-\infty}^{t} \frac{d}{dt'}\left(e^{-\frac{(t'-t_\mu)^2}{2t_\sigma^2}}\right)e^{\lambda t'} dt' = \\
= -\tau/\alpha+\gamma H_0 e^{-\lambda t}\int_{-\infty}^{t} \frac{t_\mu-t'}{t_\sigma^2}e^{-\frac{(t'-t_\mu)^2}{2t_\sigma^2}}e^{\lambda t'} dt'.
\end{split}
\label{eq13}
\end{equation}

The integral above cannot be expressed using elementary functions. Nevertheless, it can be calculated numerically with any given accuracy. 

\begin{figure}
\centering
\includegraphics[width=9cm]{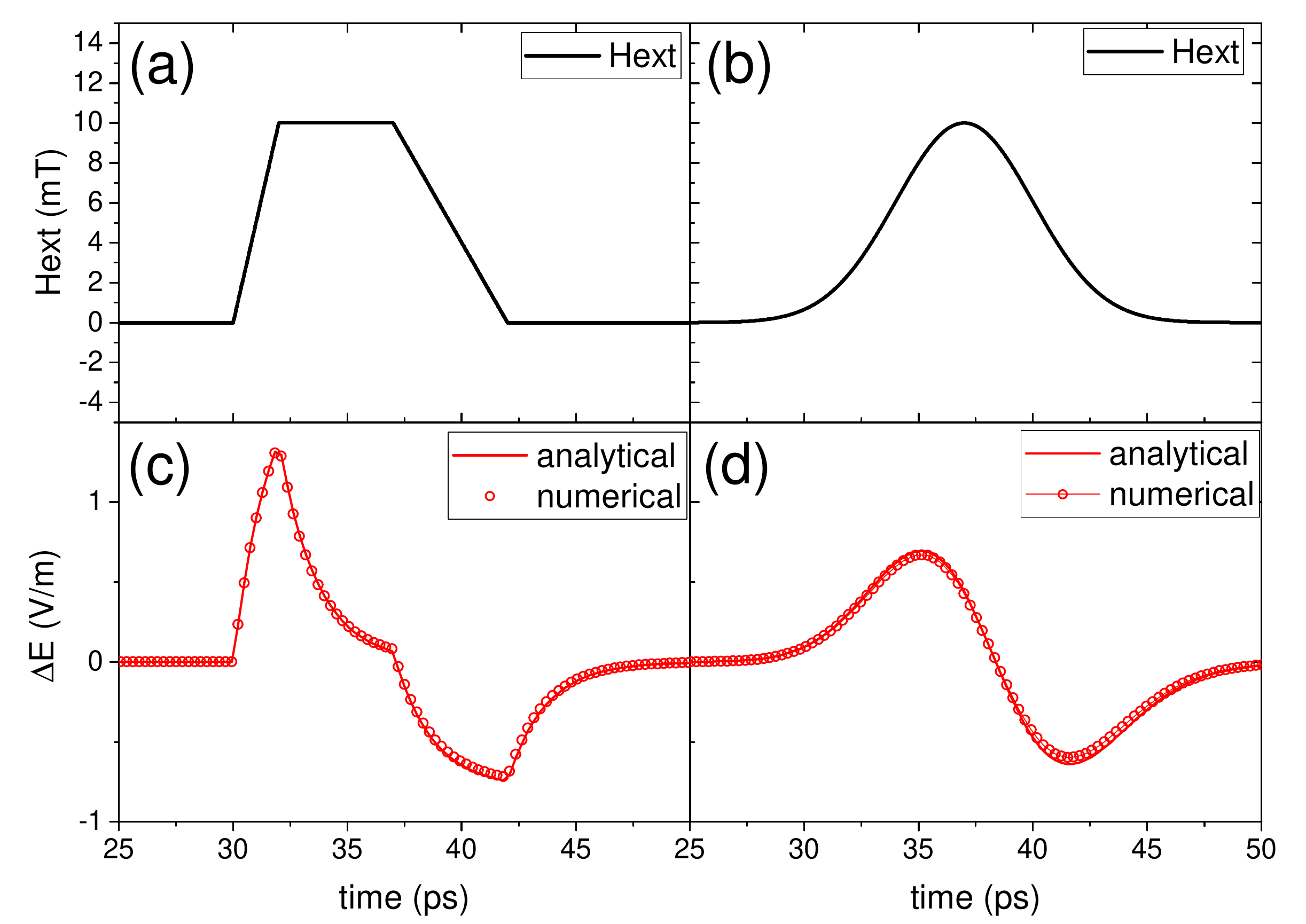}
\caption{Magnetic field pulses of trapezoidal (a) and Gaussian-like (b) shape and the respective antiferromagnetic oscillator responses to their presence (c) and (d). For the trapezoidal shape, we used slopes $\beta_1$ = 5 mT/ps and $\beta_2$ = -2 mT/ps. For the Gaussian-like shape, we used parameters $t_\mu$ = 37 ps and $t_\sigma$ = 3 ps.}
\label{fig-pulses}
\end{figure}

To illustrate and verify the solutions derived above, we conducted simulations of the antiferromagnetic oscillator system described by LLGS equations \ref{LLG1} and \ref{LLG2} . The parameters of the system were the same as in the previous section, except for an additional magnetic field in the form of a trapezoid (see Fig. \ref{fig-pulses} (a) ) or Gaussian-like pulse (Fig. \ref{fig-pulses} (b) ). The maximum amplitude of pulses was equal to 10 mT in both cases. The response of the antiferromagnetic oscillator to this excitation can be seen in Fig. \ref{fig-pulses} (c) and (d), respectively. In both cases, the analytical solutions described by equations \ref{eq12} and \ref{eq13} were compared to the results of numerical simulations based on equations \ref{LLG1} and \ref{LLG2}. One can see that a good agreement between these two methods is achieved for the trapezoidal as well as the Gaussian-like field excitation, which indicates that the approximations we used to derive the analytical solution are valid for this kind of setup. 

\subsection{Sinusoidal magnetic field}

In the case of a sinusoidal magnetic field excitation, it is possible to derive an exact analytical expression for the oscillator response. We consider an excitation described by $H_p(t) = H_0 sin(\omega t)$, which based on Eq. \ref{general-solution} will lead to the following expressions:

\begin{equation}
\begin{split}
\dot{\phi}(t)=
-\tau/\alpha+\gamma H_0\omega e^{-\lambda t}\int_{-\infty}^t cos(\omega t') e^{\lambda t'} dt',
\end{split}
\end{equation}

\begin{equation}
\begin{split}
\dot{\phi}(t)=
-\tau/\alpha+\gamma H_0\frac{\omega}{\omega^2+\lambda^2}\left(\lambda cos(\omega t) + \omega sin(\omega t) \right),
\end{split}
\end{equation}

\begin{equation}
\begin{split}
\dot{\phi}(t)=
-\tau/\alpha+\gamma H_0\frac{\omega}{\sqrt{\omega^2+\lambda^2}}sin\left(\omega t +atan(\lambda/\omega)\right).
\label{sinusoidal}
\end{split}
\end{equation}

\begin{figure}
\centering
\includegraphics[width=9cm]{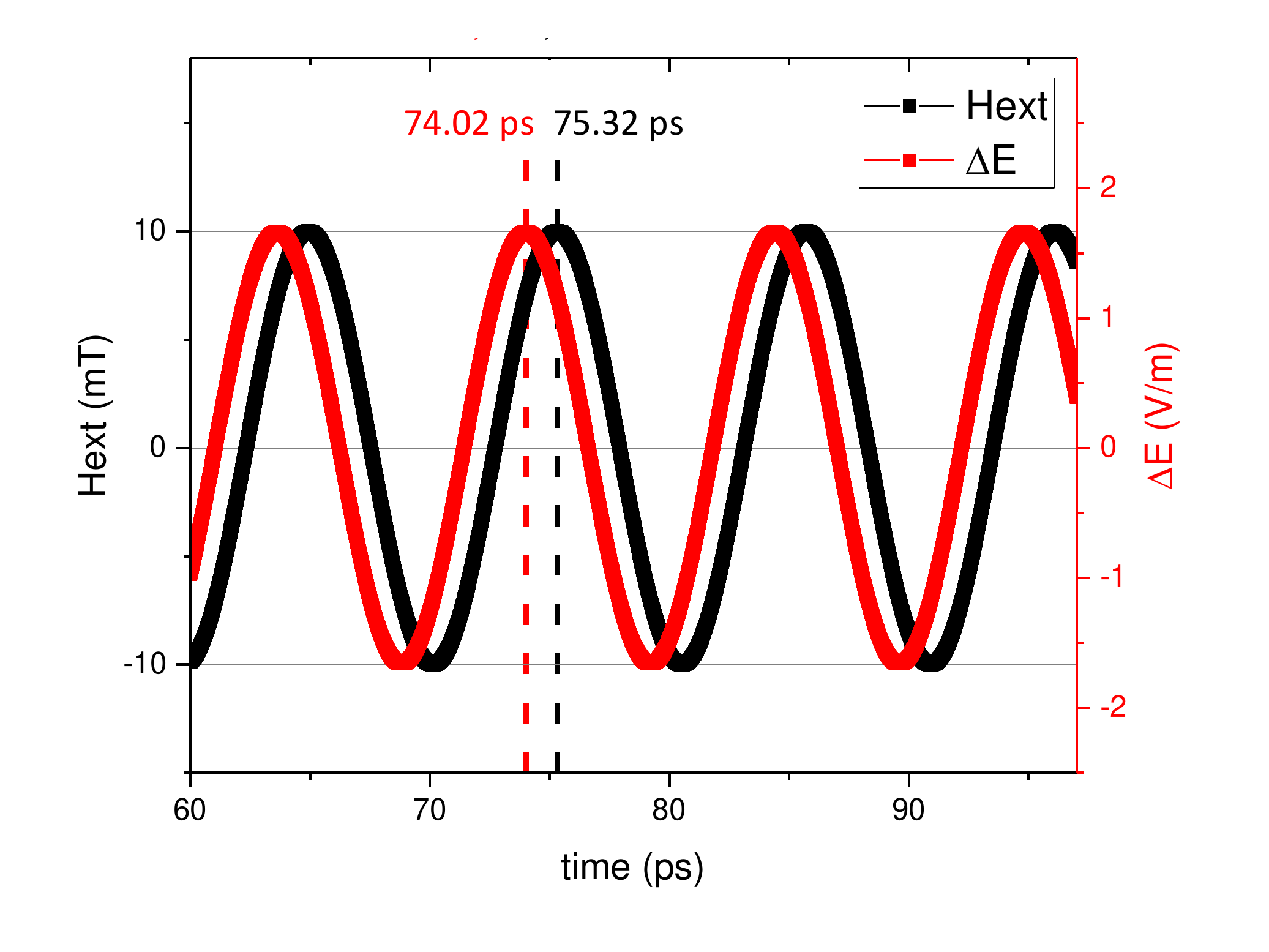}
\caption{Numerically calculated response of the antiferromagnetic oscillator (red line) to a sinusoidal magnetic field (black line) with angular frequency equal to $\lambda$. Dashed lines denote positions of example neighboring maxima.}
\label{fig-sinusoidal}
\end{figure}

One can see that the response to a sinusoidal magnetic field is also sinusoidal in character, with a phase shift factor equal to $atan(\lambda/\omega)$ included. To illustrate this, we conducted a numerical simulation of the example system described in the previous section, where $\lambda$\nobreakspace =\nobreakspace $\omega_{ex}\alpha$\nobreakspace $\approx$\nobreakspace 604.8 GHz and $\mu_0 H_0$ = 10 mT. The external field angular frequency was chosen the same as $\lambda$ (frequency f $\approx$ 96 GHz), so that the expected phase shift should be equal to 45 degrees. The result is presented in Fig. \ref{fig-sinusoidal}, where both the external field (black line) and the oscillator response (red line) are shown. We choose a pair of example neighboring maxima and determine their positions (dashed lines) as 74.02 ps and 75.32 ps. The difference, 1.30 ps, corresponds to 45 degrees phase shift in a sinusoidal signal with angular frequency equal to $\lambda$. The amplitude of the electric field changes was found to be approximately 1.7 V/m, which also is in agreement with equation \ref{sinusoidal}.

\begin{figure}
\centering
\includegraphics[width=8cm]{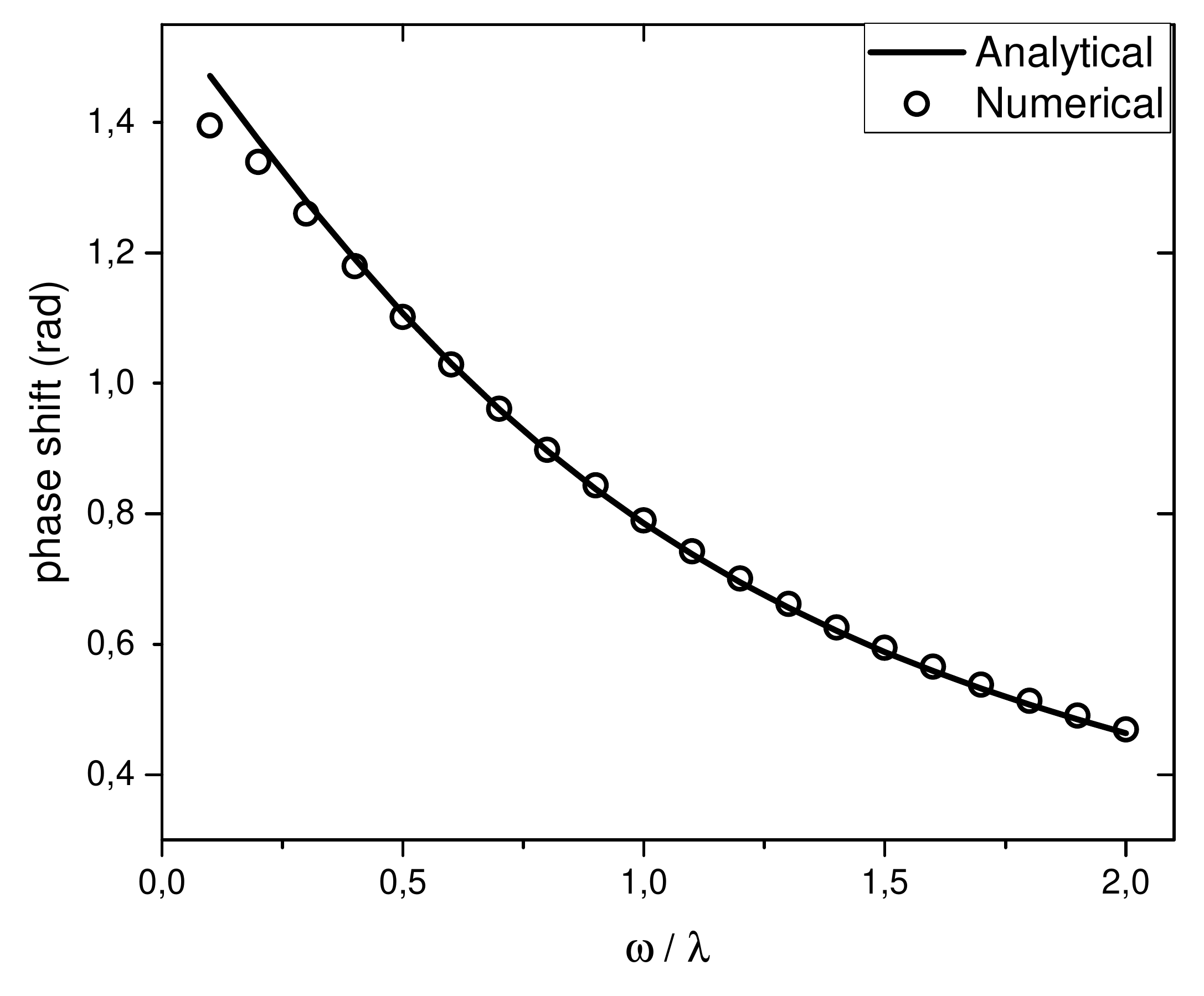}
\caption{Phase shift between the sinusoidal excitation and the oscillator response as a function of normalized angular frequency, calculated from Eq. \ref{sinusoidal} (solid line) or numerically (empty circles).}
\label{fig-phase-shift}
\end{figure}

We performed a set of numerical simulations for different angular frequency $\omega$ values, as well. Figure \ref{fig-phase-shift} presents the phase shifts recorded from simulations (empty circles) and from the analytical calculation (solid line). As the angular frequency $\omega$ is expressed in $\lambda$ units, the obtained dependence curve has a universal character and does not depend on the oscillator parameters other than $\lambda$. Applying sinusoidal magnetic fields of different frequencies may be thus utilized as a method of investigating the value of $\lambda = \omega_{ex}\alpha$, especially in a realistic antiferromagnetic oscillator system where the effective damping can be modified by the presence of spin pumping effects \cite{tserkovnyak2005nonlocal,nakayama2012geometry,khymyn2017antiferromagnetic}. Because of small absolute values of $\alpha$, the frequencies of the external signal used for that can be much smaller than the characteristic exchange frequency $\omega_{ex}$ and may not exceed tens of GHz for a typical set of parameters, making them easier to realize experimentally \cite{bonetti2009spin,braganca2013zero}.

An interesting special case of interaction with a sinusoidal magnetic field can occur when an antiferromagnetic oscillator is used to generate a THz AC signal $\dot{\phi}_{0}\times sin(\omega_{gen} t)$ due to the presence of a weak additional anisotropy \cite{khymyn2017antiferromagnetic}. If the emitted signal, originally arising purely thanks to the DC current flowing through the neighboring heavy metal layer, is allowed to generate an additional Oersted field of THz frequency, and if this field is directed fully or partially along the antiferromagnet hard axis, then according to Eq. \ref{sinusoidal} it will produce a new signal contribution of the same frequency, but different phase depending on the $\lambda / \omega$ ratio. We note here that the term derived in the Eq. \ref{general-solution}, describing the effects of interaction with an external magnetic field, can be added to the solution of the Eq. \ref{phieq} independently from the term introduced in Ref. \cite{khymyn2017antiferromagnetic} which describes the THz signal generation in the presence of a weak easy plane anisotropy. For a given set of system parameters, the generated Oersted field should be linearly proportional to the angular velocity: $\mu_0 H_{Oe}=\chi\dot{\phi}$, where $\chi$ is a proportionality constant. The effective signal emitted by the self-interacting oscillator $\dot{\phi}_{eff}$ will be then given by:

\begin{equation}
\begin{split}
\dot{\phi}_{tot}(t)=\dot{\phi}_{0} \times \sum_{n=0}^{\infty}\left(\gamma\chi\frac{\omega_{gen}}{\sqrt{\omega_{gen}^2+\lambda^2}}\right)^n \times \\
 sin\left(\omega_{gen} t +n\times atan(\lambda/\omega_{gen})\right).
\label{oscillator-fixed1}
\end{split}
\end{equation}

For sufficiently small $\chi$ and for generation angular frequencies $\omega_{gen}$ significantly larger than $\lambda$, the expression above can be approximated by:

\begin{equation}
\begin{split}
\dot{\phi}_{eff}(t)\approx  \frac{\dot{\phi}_{0}}{1-\zeta} \times sin\left(\omega_{gen} t + atan \left(\frac{\lambda\zeta}{\omega_{gen}(1-\zeta)}\right)\right),
\label{oscillator-fixed}
\end{split}
\end{equation}

where we put $\zeta = \gamma\chi\omega_{gen}/\sqrt{\left(\omega_{gen}^2+\lambda^2\right)}$. One can see that the effects of oscillator self-interaction conveyed by the Oersted field lead to a modification of the THz signal amplitude and to introduction of an additional phase shift, but do not affect the generation frequency.

\begin{figure}
\centering
\includegraphics[width=9cm]{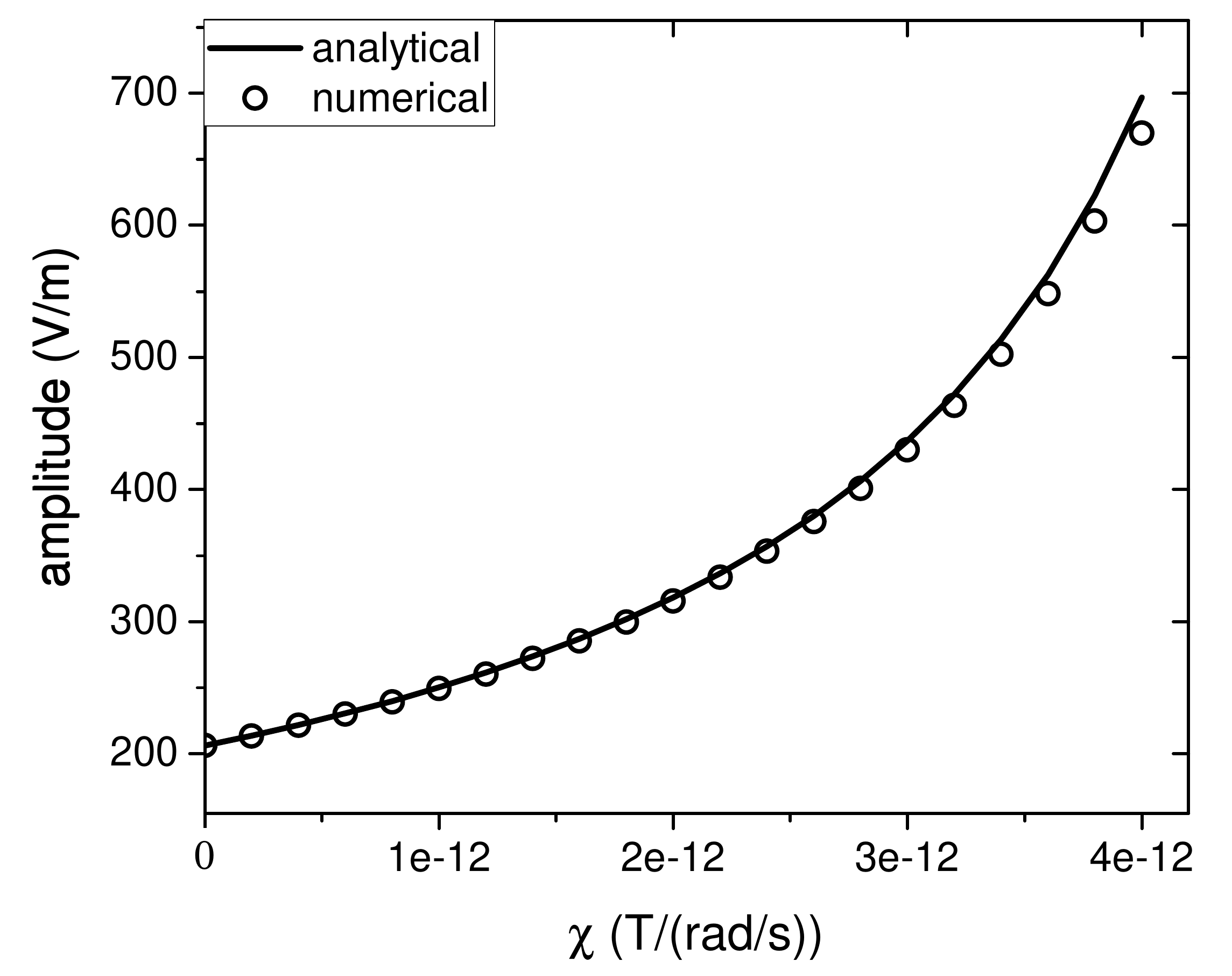}
\caption{Amplitude of the oscillator AC signal as a function of proportionality constant $\chi$ between $\dot{\phi}$ and generated Oersted field, calculated analytically from Eq. \ref{oscillator-fixed} (solid line) and numerically (empty circles).}
\label{fig-self}
\end{figure}

Figure \ref{fig-self} presents amplitude of the THz signal amplitude as a function of $\chi$ calculated using the expression above (solid line) and numerically from equations \ref{LLG1} and \ref{LLG2}, again for a system identical with the one described previously except for the presence of additional coupling mechanism between $\chi$ and Oersted field. The agreement between analytical and numerical solutions deteriorates with increasing $\chi$, which is consistent with approximations made to derive Eq. \ref{oscillator-fixed}. One can also see that the presence of the Oersted field coupled to $\dot{\phi}$ oscillations works as a positive feedback mechanism which increases the amplitude of the signal. In a device geometry where the coupling mechanism is sufficiently strong, this increase could be an important factor improving the oscillator generation power. On the other hand, it would likely require a setup where the final signal is detected as current rather than voltage THz changes, potentially limiting its applicability. 

The equation \ref{oscillator-fixed1} can be generalized further to include a presence of an additional phase shift between the generated output and the applied Oersted field 
$\Delta\phi$:

\begin{equation}
\begin{split}
\dot{\phi}_{tot}(t)=\dot{\phi}_{0} \times \sum_{n=0}^{\infty}\zeta^n \times \\
sin\left(\omega_{gen} t +n\times\left(atan\left(\lambda/\omega_{gen}\right)+\Delta\phi\right)\right). 
\label{oscillator-fixed-shift}
\end{split}
\end{equation}

\begin{figure}
\centering
\includegraphics[width=9cm]{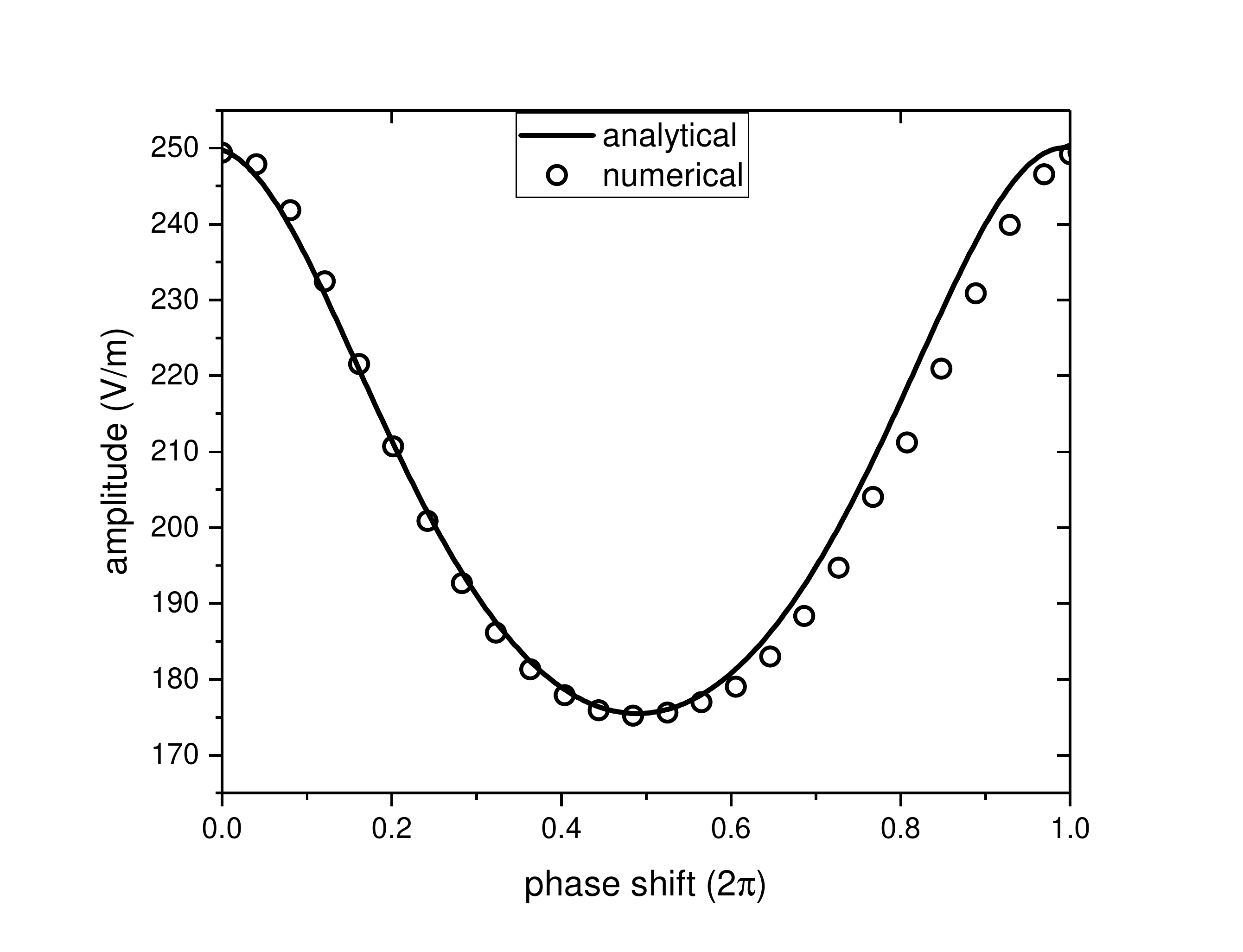}
\caption{Amplitude of the oscillator AC signal as a function of additional phase shift $\Delta\phi$ between the output signal and the generated Oersted field, calculated from analytical Eq. 19 (solid line) and numerically from equations 1 and 2 (empty circles).}
\label{fig-revise}
\end{figure}

Figure \ref{fig-revise} presents the amplitude of the output signal as a function of $\Delta\phi$ for coupling constant $\chi$ equal to $\mathrm{10^{-12}}$ (T/(rad/s)), calculated from the analytical expression above as well as from numerical integration of equations \ref{LLG1} and \ref{LLG2}. Other parameters remained the same as in the previous simulations. One can see that the signal amplitude is maximal if no additional phase shift is present and minimal for phase shift around $\pi$. This behavior is qualitatively different from the one expected for ferromagnetic spin-torque oscillators \cite{zhou2007intrinsic,zhou2008tunable}, which is explained by the fact that, in our case, the term responsible for signal generation and the term responsible for interaction with field do not depend on each other. We note here that the Eq. \ref{oscillator-fixed-shift} could be interpreted more broadly as a description of not only a self-interacting oscillator, but also of two oscillators interacting with each other. In such a setup, magnetic field phase effects \cite{zhou2007intrinsic} mentioned previously are likely to play an important role, especially if the system is further generalized towards a network of interacting oscillators. 

\subsection{Sequence of magnetic field steps}

In section \ref{sec2} we have shown that an antiferromagnetic oscillator responds to abrupt changes of an external magnetic field in a predictable manner. Thanks to the nature of the spin pumping effect, which is dependent on the magnetization vectors time derivative, such a response can be produced very rapidly. We consider now a system where a sequence of magnetic field steps of negative or positive values was applied to the oscillator. Effectively, this corresponds to a magnetic field changing its direction repeatedly, similarly to e.g. a field generated by bits on a magnetic hard disk drive (HDD) \cite{braganca2010nanoscale}\cite{checinski2017spin}. Unlike HDD read head devices, though, in the case of the antiferromagnetic oscillator the characteristic timescale of the field changes can be in the range of picoseconds, which we believe may be interesting from the applications in THz technology point of view. 

\begin{figure}
\centering
\includegraphics[width=9cm]{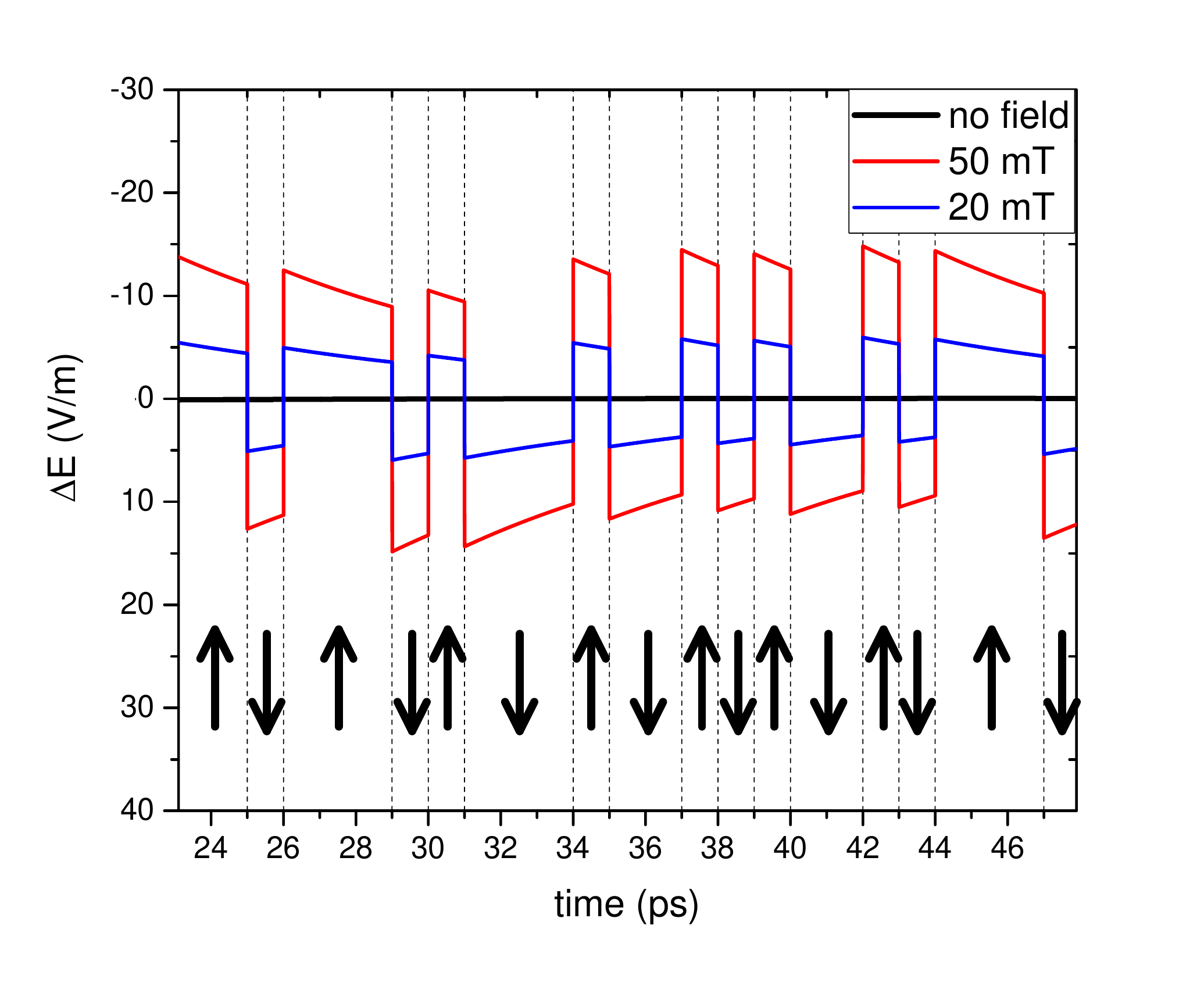}
\caption{Numerically simulated response of the antiferromagnetic oscillator to a sequence of 20 mT amplitude field steps (blue line) and 50 mT amplitude field steps (red line). The black line shows the reference level which we obtained for no external field applied. The black arrows denote the direction of the external field and the dashed lines denote when the consecutive steps occurred.}
\label{fig4}
\end{figure}

Figure \ref{fig4} presents the results of a numerical simulation where the system described in the previous sections was subjected to a sequence of positive and negative magnetic fields (see black arrows in Fig. \ref{fig4}). The time intervals between field direction changes, depicted as dashed black lines, were random integer multiples of 1 ps. Other calculation parameters were the same as described previously. The figure shows the oscillator output as a function of time in the case of no external field (black line), a sequence of 20 mT amplitude (blue line) and a sequence of 50 mT amplitude (red line). The outputs consist of superpositions of reponses to a single magnetic field step presented in section \ref{sec2}. One can see that the obtained oscillator signal retains the character of the original sequence.

\section{Summary}

We developed a comprehensive theoretical approach to the problem of an antiferromagnetic oscillator interacting with external magnetic fields. Starting from the framework based on LLGS equations and the assumption of rotation taking place in the easy antiferromagnetic plane only, we proposed a generalized form of the solution to the equation of oscillator dynamics and verified it using numerical simulations. Based on this result, we derived the formula for the response to an external field of any given shape and described several typical cases of the interactions with the external field, including a trapezoid and a Gaussian pulse, a sinusoidal field and a sequence of field steps. In the case of a sinusoidal field, we found that the inherent dynamics of the oscillator leads to a presence of a phase shift in the output signal, which could be utilized to investigate device parameters, such as effective damping, with excitations in the range of GHz only. We also described how an oscillator working as a THz emitter device, which has been proposed recently, could interact with the Oersted field produced by its own output signal, which would increase the effective amplitude and power of THz oscillations. Finally, we performed a simulation of an oscillator reading a picosecond sequence of magnetic field steps, which can be interesting in the context of future development of THz technology devices.

\section*{Acknowledgements}
The authors would like to express their gratitude to dr. Ivan Lisenkov and dr. Roman Khymyn for their critical remarks and fruitful discussion. We acknowledge the grant Preludium 2015/17/N/ST7/ 03749 by National Science Center, Poland. T.S. acknowledges the grant SpinOrbitronics 2016/23/B/ST3/01430 by National Science Center, Poland. Numerical calculations were supported by PL-GRID infrastructure.



\bibliography{bibliography}

\end{document}